\documentstyle[epsf,preprint,aps]{revtex}
\begin{document}

\title{Hiking in the energy landscape in sequence space: a bumpy road
to good folders.}

\author{G. Tiana$^{1}$,  R. A. Broglia$^{1,2,3}$, E. I. Shakhnovich$^4$}
\address{$^1$The Niels Bohr Institute, University of Copenhagen, 
Bledgamsvej 17, 2100
Copenaghen, Denmark.}
\address{$^2$Dipartimento di Fisica, Universit\`a di Milano,
Via Celoria 16, I-20133 Milano, Italy.}
\address{$^3$INFN, Sez. di Milano, via Celoria 16, I-20133 Milano, Italy}
\address{$^4$Department of Chemistry, Harvard University, 12 Oxford Street,
Cambridge, MA 02138}

\date{May 19, 1999}

\maketitle

\bigskip

\newpage

\begin{abstract}
With the help of a simple $20$ letters, lattice model of heteropolymers, 
we investigate the 
energy landscape in the space of designed good-folder sequences. 
Low-energy sequences 
form clusters, interconnected via neutral networks, 
in the space of sequences. 
Residues which play a key role in the foldability of the chain and in the
stability of the native state are highly conserved, even among the chains
belonging to different clusters.
If, according to the interaction matrix,
some strong attractive interactions are almost degenerate (i.e.
they can be realized by more than one type of aminoacid contacts)
sequence
clusters group into a few super-clusters. Sequences belonging to different
super-clusters are dissimilar, displaying very small ($\approx 10\%$)
similarity, and residues in key-sites
are, as a rule, not conserved. 
Similar behavior is observed in the 
 analysis of real protein sequences.
\end{abstract}
\newpage

\section{Introduction}

The degeneracy of protein folding code is 
well-documented: many sequences exist that can fold to similar
native conformations \cite{BT98,Chothia:97}. Besides homologs, i.e. proteins
that have a clear evolutionary connection and homologous
sequences (differing by a few neutral mutations) which  are often
(but not always) functionally related, there exist analogs,
i.e. structurally similar proteins that 
have non-homologous sequences (with sequence 
identity less than 20\%), unrelated
functions and no evident evolutionary relation \cite{Chothia:97,ROST_FD}.
The analysis of the origin of analogs emphasizes the physical
aspect of molecular evolution because they share common fold
but not function.

An important question is whether present analogs emerged as  
a result
of a long divergent evolution or they originated from
dissimilar sequences/structures and converged to
structurally homologous folds? A physical approach
to address this question is to study the ''topography'' of the space of  
sequences which  
fold
into the same target conformation. In particular, the connectivity of the
space of sequences via neutral nets (i.e. single mutations that preserve the
foldability into the "native" structure \cite{GG:97}) may be a good evidence for 
divergent evolution as an origin of analogs, while 
the presence of disconnected ''isles'' in sequence space
would be an argument in favor of the convergent evolution origin
of analogs.

The aim of the present work is to address this question rigorously 
by analyzing 
the properties of the sequence space in an exactly tractable lattice model
of a protein.
For this purpose, we 
use a simple $20$--letters 3-dimensional lattice model of heteropolymers
\cite{CHPH,SSK1} and contact
energies obtained from the statistical analysis of real proteins \cite{MJ}.
Good--folder sequences are characterized by a large gap
$\delta$ (as compared to the
standard deviation of the contact energies $\sigma$),
 between the energy of the sequence in the native 
conformation and the lowest energy (threshold $E_c$) of the 
conformations structurally dissimilar to the native one \cite{GSW,SSK1}.
In other words,
good folders are associated with an "order parameter" $\xi=\delta/\sigma \gg 1$
(this quantity is closely related to the z-score \cite{EISENBERG}).
Because the threshold  energy $E_c$ entering in the definition of the gap
depends 
only on the 
composition of the chain \cite{PRLF}, 
the knowledge of the energy landscape of the
space of sequences with fixed amino acid composition enables us to 
select good folders and to investigate their properties. In particular,
it is possible to study the space of folding--sequences within the context of
ground state properties, using only equilibrium thermodynamics.

The use of the lattice model to study properties of energy landscape
in sequence space has two clear advantages. First,
folding properties of each sequence can be examined 
directly from folding dynamics 
simulations which are quite feasible for such models.
Second all sequences designed to fold into a given conformation
have the same length. This factor simplifies
comparison between sequences
eliminating the need to  introduce 
of arbitrary gap/insertion penalties.

In particular, a single parameter, 
defined as the ratio of non-matching residues to
the total number of residues (an overlap parameter) 
in a straightforward
alignment of two sequences 
  can serve as a natural measure of similarity between them.
Using this parameter as a  metric in the sequence
space we show that sequences folding to the same native conformation
cluster together in a rather articulated, 
hierarchical fashion, in the space of
sequences. Such clustering has a direct evolutionary implications
which we discuss at the end of this paper. 

The model used in this  calculation is a standard lattice model
that represents amino acids as
point-like beads on the vertices of a cubic lattice, connected by
rigid links (covalent bonds). Different monomers interact 
through the nearest-neighbor (in space) Hamiltonian
\begin{equation}
H\left(\{r_i\},\{\sigma_i\}\right)=\sum_{ij} B_{\sigma_i \sigma_j} \Delta(r_i -
r_j),
\label{H}
\end{equation}
where $\sigma_i$ and $r_i$ are the kind and position of the $i$th residue,
and $\Delta(r_i - r_j)$ is $1$ if $r_i$ and $r_j$ are 
non--consecutive nearest--neighbors, and zero otherwise. 
The parameters $B_{\sigma_i\sigma_j}$ are the contact energies between the 
amino acids.


As noted above, the similarity
parameter $q_s^{\alpha\beta}$
 between sequence $\alpha$
and sequence $\beta$, defined as
\begin{equation}
q_s^{\alpha\beta} = \frac{1}{N}\sum_{i=1}^N
\delta(\sigma^{\alpha}_i,\sigma^{\beta}_i),
\label{qs}
\end{equation}
serves as a metric in sequence space.
Here $\delta(\sigma^{\alpha}_i,\sigma^{\beta}_i)$ is $1$ if the $i$--th
residue of sequence $\alpha$ is equal to the $i$--th residue of
sequence $\beta$ and zero otherwise. 
A natural parameter to describe energy 
surface in sequence space is
the
distribution
\begin{equation}
P(q_s)=<\delta(q_s-q_s^{\alpha\beta})>,
\label{pq}
\end{equation}
where $<>$ indicates a proper (''thermal'') average over all possible pairs of
sequences $\alpha$ and $\beta$.

To study the properties of the distribution $P(q_s)$ we have chosen
the $48$-mer target structure (native conformation) shown in the inset
to Fig. 1. 
Using standard Monte-Carlo optimization in sequence
space \cite{PNAS,PE,PRLF} we designed many sequences
that have low energy in the 
target structure shown in Fig.1. The optimization  
in sequence space was carried out with constant aminoacid
composition corresponding to the ''average'' composition of natural
proteins \cite{CREIGHT}. This condition implies optimization
of the sequence order parameter $\xi$ via optimization of the
gap $\delta$ keeping fixed the variance of interaction energies $\sigma$
(that depends on aminoacid composition only \cite{PRLF}).
The elementary MC move in sequence space that preserves aminoacid composition
is swap of two randomly chosen residues.
The acceptance of a ''move'' is controlled by applying a standard
Metropolis criterion with ''selective temperature'' $T$ that 
serves as a measure of a degree of evolutionary pressure towards
sequences having larger energy gaps. 
The contact energies
$B_{\sigma_i\sigma_j}$ were taken from Table 5 of ref. \cite{MJ} 
($\overline{B}=0$, $\sigma=0.3$).
Folding of a a typical sequence generated at $T=0.01$
(denoted $S_{48}$, cf. caption Fig. 1) was studied using standard
lattice Monte-Carlo folding simulations \cite{DEUTCH,SSK1,PRLF,PNAS}
and the thermodynamic and kinetic parameters for that 
sequence were determined. In the units we are
considering ($RT_{room}=0.6$ kcal/mol), the energy of S$_{48}$
in its native conformation is $E_{nat}=-26.97$, while the value
$E_c$ of the lowest dissimilar conformation is $E_c=-21.50$, leading
to a gap $\delta=5.47$ and to an "order parameter" $\xi=18.2$. We are thus
guaranteed that S$_{48}$ folds fast \cite{SSK1,PNAS,Broglia_99}. 
Indeed, its mean first passage
time to the native  conformation 
is $3.3 \cdot 10^6$ MC steps at the optimal folding temperature
($T_f=0.28$).
Also, any sequence S$'_{48}$ obtained from mutations which do
not alter significantly the chain composition 
and which, in the native conformation, has an
energy smaller than $E_c$, (i.e. an excitation energy lying within the
gap $\delta$) will fold \cite{Tiana_98}. A number
of other designed sequences  were
studied with similar results for folding thermodynamics and kinetics. 

The sequence design procedure was performed at different selective 
temperatures
$T$, recording each time $1000$ sequences with energy lower
than a given threshold $E_{th}=E_c$ and separated by  $1000$ MC
steps. From the collected sequences it was possible to calculate
the distribution $P(q_s)$ at different temperatures. 
The results are shown in Fig. 1 for several simulation temperatures ranging 
from $T=0.01$
to $T=0.08$.  
The $P(q_s)$ plot 
recorded at $T=0.01$ features a number of, peaks the 
most pronounced being centered at $q_s=0.1$, $q_s=0.55$ and  $q_s=0.95$.
The peak at $q \approx 0.55$ 
is not present in the $P(q_s)$ plots associated with higher 
evolutionary temperatures.
For example, at $T=0.08$, only two peaks are apparent, one at $q=0.1$ 
and a set of smaller peaks in the high-q region.

Since our MC design sampling probes different regions of sequence
space at low-- and at high--temperatures,
the comparison of the corresponding results allows to get insight into
the nature of the energy landscape in sequence space in a wide range of energies.
Apparently, the low-temperature simulation ($T=0.01$) 
samples the bottom part of the energy spectrum in sequence space.
The peaks of $P(q_s)$ at $q_s\approx 1$ and at $q_s=0.55$ indicate that the
low-energy part of space of designed sequences can be divided into
clusters, each cluster containing sequences that 
differ 
only by a few mutations. 
Furthermore, within each cluster, there are $6-8$ sites  in which
the associated residues are $100 \%$ conserved. These sites are the ones denoted
as
"hot" in the ref. \cite{Tiana_98}, controlling both the folding 
and the aggregation \cite{aggreg} of proteins.
The size of each cluster varies to a 
considerable extent but no sequence is isolated.
This means that, for each sequence, it is possible to introduce a swapping
of amino acids between some sites and still obtain a good folding sequence
belonging to the same cluster. 
The maximum
number of sequence changes that lead again to a sequence  
belonging to a cluster is about $15 \%$ of the length of the chain.
If this limit is exceeded, a wider cooperative rearrangement of
 residues (between $40\%$ and $60 \%$, corresponding to the second peak in
Fig. 1) is needed in order to produce a 
sequence which again displays an energy within 
the gap $\delta$ in the native 
conformation. Such greater sequence 
rearrangements correspond
to moves 
in the space of sequences, from one cluster to the next one.

Since all sequences belonging to a cluster are equivalent,
i.e. they display $>90\%$ overlap among them, it is
possible to compare different clusters by comparing sequences
chosen at random from them. In other words, each cluster
behaves like a class of equivalence. 
From a detailed
analysis of the sequences belonging to different clusters, it turns
out that clusters group into "super-clusters". Clusters within
each super-cluster are associated with a similarity parameter 
$q_s\approx 0.55$, while clusters belonging to different super-clusters 
have $q_s\approx 0.1$.

Within each super-cluster, ``hot'' sites are in the same position and the
type of residues occupying each of these sites is conserved in $100 \%$ of
the sequences. This fact emphasizes once more the importance of ``hot''
sites to govern the low-energy properties of protein sequences. 
Furthermore, half of the ``hot'' sites of sequences belonging to the
first super--cluster are in the same positions as the
``hot'' sites of sequences of the second super--cluster and the
other half move to neighboring sites.
The common ``hot'' sites are different for residues which are very
similar (meaning that the columns of MJ matrix indicating
their interaction with all the other monomers, are very similar, i.e. D with E 
and Q, K with R). This suggests that the presence of the two
super-clusters may be due to the quasi-degeneracy of these matrix
elements. The substitution of a monomer in a ``hot'' site with a very
similar one causes a small rearrangement of the other
``hot'' sites, and to a much larger degree the rearrangement of the other
("cold") residues.
To substantiate this scenario, we have repeated the calculation
with a random-generated matrix having the same distribution of elements
of the MJ matrix. Because in this case all the columns of the matrix are
different, we expect no super--clusters. In fact, the distribution 
$P(q_s)$ for this case is 
very similar to the distribution shown in Fig. 1, except for the absence 
of the peak at $q_s\approx 0.1$ (data not shown).

The schematic illustration of the low energy part
of the sequence space  is sketched in Fig.
2A. For each good folder sequence it is possible to mutate
a few ``cold'' sites, up to $15 \%$ of them and still obtain a good
folding sequence. In this way one produces a cluster. If more ``cold'' sites
are changed, there is a partial rearrangement of many other ``cold''
sites to keep the energy low, leading to
different clusters. When a ``hot'' site is 
mutated with a similarly
strongly interacting
residue (if the interaction matrix 
allows for such a possibility), 
the other
``hot'' sites become slightly 
rearranged, while the ``cold'' sites
are heavily rearranged, the net 
result being a new super-cluster.

An important aspect of this analysis concerns the
 height of the ''barriers'' in the sequence space.
This question is closely related to the issue of connectivity
of clusters via a neutral network, i.e. a  set of mutations that preserve
the ability of all intermediate sequences to fold into the
same structure. While the question of what constitutes a
viable folding sequence is a complicated one having a number
of aspects (thermodynamic, kinetic and functional 
\cite{EVOL2,GG:98,GG:97} a necessary condition for a sequence
to fold into a given structure is that its energy in this structure
is lower than $E_c$ \cite{PRLF,PNAS,GSW}. From this ''minimalistic''
point of view, clusters are connected by neutral networks
while superclusters are not - the top of the barrier
in sequence space between them lies above the level of $E_c$, i.e.
the muation ''pathway'' between superclusters passes through the ''sea''
of non-folding sequences.
This is schematically illustrated in Fig. 2A. 

The schematic representation of 
the structure and connectivity of the part of the sequence space sampled
at higher temperatures (i.e. that fits the native structure with higher
energy but still lower than $E_c$) is shown in Fig.2B. The main difference
between $T=0.01$ and higher temperature simulations is that
the peak of $P(q_s)$ at $q_s=0.55$ does not exist at 
the high temperature samplings.
This result indicates that the structure of the sequence space 
sampled at $T=0.08$
does not include clusters. However the superclusters
remain the same with the same ''hot'' sites conserved in each of them.
Fig. 2B(c) schematically represents the energy landscape
in the sequence space obtained  both from low temperature sampling
and high temperature sampling. It clearly shows that 
there are two characteristic barriers between clusters and between
superclusters. When the sampling temperature exceeds the former the
structures are not observed and the simulations samples
broad basins each corresponding to a supercluster, in our terminology.

We also note that the actual histogram of $P(q_s)$ at low temperature
(T=0.01)
depends on the MC-time step with which the space is probed.  In the analysis
presented above it was sampled with 1000 MC step interval. Apparently
the characteristic number of mutations needed to cross barrier 
between clusters
is greater than 1000 and consequently the peak at of $P_(q_s)$
at $q=0.55$ is observed. However, increasing the sampling step to 10000
(again at $T=0.01$) leads to the histogram of $P(q_s)$ that is qualitatively
similar to that associated with higher temperature. This is fully consistent
with the above picture: the typical number of mutations needed 
to switch between clusters
at $T=0.01$ is less than 10000 steps hence the fine picture of clusters
disappears at that sampling step. Thus temperature and sampling interval
appear to be 
 complementary tools to sample the complex energy landscape in sequence
space.
To proceed further, the universality of these findings has to be 
tested against the conformation chosen as native and to 
the interaction matrix.

We have analyzed the distribution $P(q_s)$ for ten different compact
conformations of a $48$ monomers chain.  In all cases the distribution 
$P(q_s)$ taken at low temperature $T=0.01$ with sampling interval of 
1000 mutations
exhibits three peaks, 
one 
of them is centered at $q_s\approx 0.9$. The two peaks observed at lower
values of $q_s$ are always well separated
from this peak by a minimum at $q_s\approx 0.7$. The position
and width of the low--$q_s$ peaks seem to be somewhat dependent on the specific
conformation. This is in agreement with the interpretation we gave
of the existence of these peaks, namely that they are caused by rearrangements
of a substantial part of the sequence. The number of amino acids involved in
these rearrangements can
depend on the details of the geometry of the system and thus the degree of
similarity that the mutated sequences will display with each other.

To study the dependence of $P(q_s)$ on the interaction matrix, 
we repeated the same
calculations with a random--generated matrix (see above) and with
the matrix published by Kolinski {\it et al.} 
in ref. \cite{KGS}. Unlike
the MJ matrix, those of ref. \cite{KGS} do not display any 
similar columns. Consequently,
the associated histogram of
$P(q_s)$ shows only two peaks. They are located 
at $q_s=0.55$ and $q_s=0.95$ when simulations are run at $T=0.01$ and
sampling each 1000 mutations. At higher temperature or longer sampling step 
the ''left''
peak moves to low--$q_s$ values of $q_s$ ($q_s \approx 0.2$). 
Furthermore, in this case  for each structure, the type
of residue occupying  ``hot'' sites is unique because with the
set of parameters of ref. \cite{KGS} only one type of interactions (hydrophobic)
can provides substantial stabilization to the hot sites.
Hence, in this case, superclusters are not observed in the
 energy landscape in sequence space. 

In the case of models containing only two kinds of residues (Hydrophobic and
Polar), 
the assertion that $E_c$, the lowest  boundary energy of states
that are structurally 
different from the native one depends only on the composition of the 
chain, may not hold \cite{SGS,DK}. Therefore, in this case, 
it may not be possible to describe
in a simple fashion the space of large-gap sequences in terms of
ground-state properties. Nonetheless, it is
still necessary (but now not sufficient) for 
designed sequences to have, in
the native conformation, 
low energy as compared, for 
example, to the average energy of compact configurations. Calculations
performed with the matrix elements used in ref. \cite{Tang} show
that $P(q_s)$ has only one peak close to $q_s=1$.
We  also analyzed sequences composed of three kinds
of residues. The
behavior in this case is more similar to that found for $20$ kinds of
residues. 
Three seems then to be the minimum quantity of
different residues necessary to have a non-trivial shape of $P(q_s)$.

Furthermore, we repeated the calculations optimizing the sequences with
an algorithm that does not require the constraint of constant amino acid
composition of the chain. This algorithm \cite{MMES} maximizes the stability
of the sequences, approximating the free energy of 
non-native states with a high
temperature expansion. The resulting shape of $P(q_s)$ is very similar to the one
shown
in Fig. 1, except for the fact that the low-q peak is very small. This last
feature seems
to be a non physical consequence of the high temperature approximation.
 

An analysis of the distribution of the parameter $q_s$ associated 
with real proteins
was carried out by Rost in ref. \cite{ROST_FD}. The values of
$q_s$ are obtained by the comparison of pairs of sequences that fold
to similar 3D structures. Sequences and structures were 
selected from the PDB database  in a way to minimize bias
due to the finiteness of the set. 
The function $\overline{P(q_s)}$ thus obtained displays two well defined peaks
centered at $q_s=0.08$ and $q_s\approx 1$, and a continuum of smaller
peaks in between. The overline on the distribution of $q_s$ indicates
that the results of Rost have the meaning of an average of $P(q_s)$,
as defined in Eq. (\ref{qs}), over many different target structures.

The results of Rost for real sequences seem 
to match better the 
pattern of $P(q_s)$ obtained
in our simulations at low temperature ($T=0.01$) 
and at low sampling frequency of 10000 mutations or
at higher temperature ($T=0.08$).
As was already stated, the energy landscape 
in sequence space is complex
so that the way how it is sampled may 
affect the apparent observed correlations. 
Specifically, it is important to make proper
weighting of close homologs. Clearly,
exclusion or underweighting of  homologs in protein sequence 
comparison would generate a highly bimodal $P(q_s)$ distribution
similar to that presented in \cite{ROST_FD} while
inclusion of sequence homologs would give rise to peaks at
intermediate values $q_s$. Such peaks are noticeable in
the plots of $P(q_s)$, presented in \cite{ROST_FD}.
However their heights are relatively low possibly due to 
underweighting of close homologs
in \cite{ROST_FD}.

In order to achieve a better understanding of the relation between 
evolutionary
temperature used in this study and the measure of 
natural evolutionary pressure, 
we analyzed the sequence entropy per site
\cite{ALAMUT}. For each site $i$, entropy is defined as
$S(i)~=-~\sum_{\sigma=1}^{20} p_\sigma(i) \ln p_\sigma(i)$, $p_j(i)$ being the 
probability to find a residue of kind $\sigma$ in site $i$. The
histogram of sequence entropy per site for the target structure
of Fig. 1 is displayed in Fig. 3(a). At low selective 
temperature ($T=0.01$,
solid line) the histogram shows two distinct peaks, one corresponding to highly
conserved sites (11 monomers) and one corresponding to
 poorly conserved sites (37 monomers). Among the 11 highly
conserved sites there are those that determine
the "super-cluster" ("hot"
sites, see above) and participating in the partial folded intermediates
\cite{aggreg} which, by assembling together, give rise to the
folding core. At high temperature ($T=0.10$), there is only one peak, 
approaching the equiprobability behavior
($S(i)=2.99$ for each site). A transition regime between these two 
patterns  takes
place at $T=0.08$.

This peculiar bimodal distribution of sequence entropies
can be compared with real proteins.
Fig.3(b) shows the distribution of sequence entropies in structurally aligned 
families of proteins having Immunoglobulin (Ig) fold (a detailed explanation 
of how this quantity was calculated can be found in \cite{EVOL2,EVOL_JMB}).
We see a remarkable qualitative agreement between the low-temperature
($T\leq 0.08$) 
model simulation results
presented in Fig3(a) and that for the Ig family proteins: 
both show
a pronounced bimodal distribution. As explained above, for the case
of lattice sequences, the bimodal 
distribution
of individual entropies is related to the complicated structure
of the energy landscape in sequence space. The same
feature observed
in one the most populated protein families is suggestive of complex
energy landscape in sequence space of real proteins.

Summing up, we find that the space of sequences that fold
into the same native conformations
is quite rich, being partitioned into
clusters and superclusters. 
Sequences that belong to different clusters have little
similarity despite the fact that they are able to fold into
the same native conformation. Nevertheless 
sequences belonging to the same
super-cluster (but different clusters) have the same aminoacids in a few
strategic ''hot'' core positions \cite{Tiana_98,aggreg}. Finally, aminoacids in core
positions vary between superclusters.

These results bear striking similarity to recent findings
for real proteins. 
In fact, the analysis presented in \cite{EVOL2,EVOL_JMB} revealed
universally conserved core positions in several protein folds.
In some cases the analogs had the same types of aminoacids in these
positions despite the lack of sequence similarity anywhere else.
However, in several cases correlated mutations in the core
were observed which preserved strong stabilization of the
core but with different sets of aminoacids. For example, in some cases
strong aliphatic hydrophobic contact 
was replaced by disulfide bond. Another example of this cehaviour is
provided by 
aromatic rescue of glycine in $\beta$-sheets \cite{REGAN_FD}
where two aliphatic hydrophobic groups in cores of 
$\beta$-sheets
can be replaced by glycine and an aromatic group
without loss of stability. Using the terminology of the present
paper, these situations correspond
to transitions from one supercluster to another.

A non-trivial distribution of barriers
translates into complex dynamics with multiple discrete
time scales. In our case the results relate to evolutionary dynamics. 
In particular, we found that evolutionary dynamics 
may proceed with distinctly different rates for different parts
of a protein.

Our study may
provide a meaningful ''toy'' model of convergent and divergent
evolution. The analysis shows that transitions
between clusters occur over the barriers lower than
$E_c$. In other words, clusters may be connected by
neutral networks of sequences all folding into the same native conformations,
albeit with different stabilities. However, superclusters are separated
by barriers that are high enough to exceed $E_c$ at their tops (see Fig.2).
This means that it is not possible to pass from supercluster
to another supercluster via a neutral network. Correspondingly,
sequences that belong to the same cluster could have
appeared as a result of divergent evolution, while sequences
that belong to different superclusters are likely
to appear as a result of convergent evolution.
Translating this to the language of evolutionary physical
analysis of real proteins we assert that in cases
when protein analogs contain different types of aminoacids in universally
conserved cores they may have been the product of convergent evolution,
while similar aminoacids in the core may be suggestive of  divergent evolution
origin of analogs.

Another possible physical evolutionary pressure that was not taken
explicitly into account in our present calculations 
is the one towards fast folding
kinetics. 
It was argued in \cite{EVOL_JMB} that such factors as 
the need to prevent  aggregation or proteolysis may
lead to effective pressure
towards fast-folding sequences. However, it was shown in
simulations \cite{ALAMUT,EVOL2,Tiana_98,aggreg} and
for some proteins (\cite{EVOL_JMB}, J.Clarke private communication) 
that the  core residues conserved within clusters are also
the ones that participate in nucleation and thus determine the folding rate.
Therefore for such proteins, stability will be highly correlated
with folding rate and evolutionary regulation of both factors
may result in stabilization of the same set of contacts.
However it was noted in \cite{EVOL_JMB} that this is not always
the case: a counterexample is cold-shock protein family \cite{CSP2},
where stability and fast-folding are likely to be determined by
 different sets
of contacts. In these cases the evolutionary 
optimization of folding kinetics may result in
additional conserved cluster in the nucleation region.
Obviously future experimental and theoretical
studies will help to identify the physical and biological reason for
sequence conservation in structures of most protein
families.

This work has been partially supported by the Graduate
School of Biophysics, University of Copenhagen (to GT),
the INFN Sez. Milano (specific project MI31), Nato 
(Grant CRG 94023, to RAB)
and NIH (grant GM52126 to ES).

\newpage

\newpage
\section*{Figures}

{\bf FIG.1} The probability distribution for $q_s^{\alpha \beta}$ over $1000$
sequences with energy below $-25.50$ found in MC minimizations at
temperature $T=0.01$ (solid line). The target conformation is displayed in
the inset. The other curves correspond to $T=0.05$ (dash--dotted),
$T=0.08$ (dotted) and $T=0.10$ (dashed). For technical reasons, each curve
at $T>0.01$ is composed of a part calculated by recording, in the
simulation, one sequence every 1000 (left peaks) and a part calculated
picking up one sequence every 10 (right peaks). The designed sequence
$S_{48}$ (generated at $T=0.01$) mentioned in the text is
MKQVTPEWSHYAGPDCMKQGTSVGQHCNIWELENILNCYFDLSAFRIR.

\vspace{5pt}

{\bf FIG.2}
Schematic representation of the structure of the
space of sequences folding into the same conformation. The calculations
were carried out for a 48-mers chain making use of Miyazawa--Jernigan
contact energies among the amino acids. ({\bf A}) The space of 48-mers
sequences, all folding to the same native conformation (cf. (c) or (c') )
as derived from an analysis performed at very low evolutionary temperature
($T=0.01$), resulting in the distribution $P(q_s)$ shown in (a). The main
structure observed is that of superclusters (as shown in (b) and (b') in
dark grey) which are separated by potential barriers higher than $E_c$.
Within each supercluster "hot" residues (black dots in (c) and (c')  ) are
highly conserved. Between two superclusters the number, type and position
of the "hot" residues can change to some extent, the associated change in
contact energy being responsible for the large barrier shown between
supercluster (b) and (b'). A supercluster is built out of a large number
of clusters which are shown in light grey. Potential barriers among
clusters display energies all smaller than $E_c$ (cf. energy plots between
clusters inside the supercluster (b) ). \\
({\bf B}) The space of 48-mers
sequences all folding to the same native conformation (cf. (c) and (c') in
plot (A) of the figure), designed at a high evolutionary temperature
($T=0.08$). In this case the function $P(q_s)$ shown in (a) displays only
two peaks ($q_s\approx 1$ and $q_s\approx 0$).Consequently,
the space of sequences is divided only in superclusters 
((b) and (b')). In (c) we display a schematic representation of the evolution of the
space of designed sequences, making use of low-- and high--evolutionary
temperature. 

\vspace{5pt}

{\bf FIG.3}
{\bf (a)} The histogram of sequence entropy per site. At low temperature
($T=0.01$, solid line) the histogram displays two different peaks, corresponding
to separation in structure into 
high-- and low--conserved sites. At $T=0.08$ (dashed line) a
transition occurs and at higher temperatures ($T=0.10$, 
dotted--dashed line) only one peak at high entropy, showing
no particular conservation anywhere in the structure, is present.
{\bf (b)} The histogram of the distribution of sequence entropy
for proteins having Ig fold. 
Sequence entropy for a position $l$ in Ig-fold structure
(we used tenascin, 1 ten as a representative protein)
was calculated as  
$S^{across}(l)=-\sum_{i=1}^{20} p_i^{across} (l) \log p_i^{across} (l)$.
To obtain this quantity we evaluated frequencies of occurrence 
of amino acids of each type $i$
 at each position $l$ for all families ($p_i^{across} (l)$). 

\pagestyle{empty}

\pagebreak
\epsfysize 7 in
\epsffile{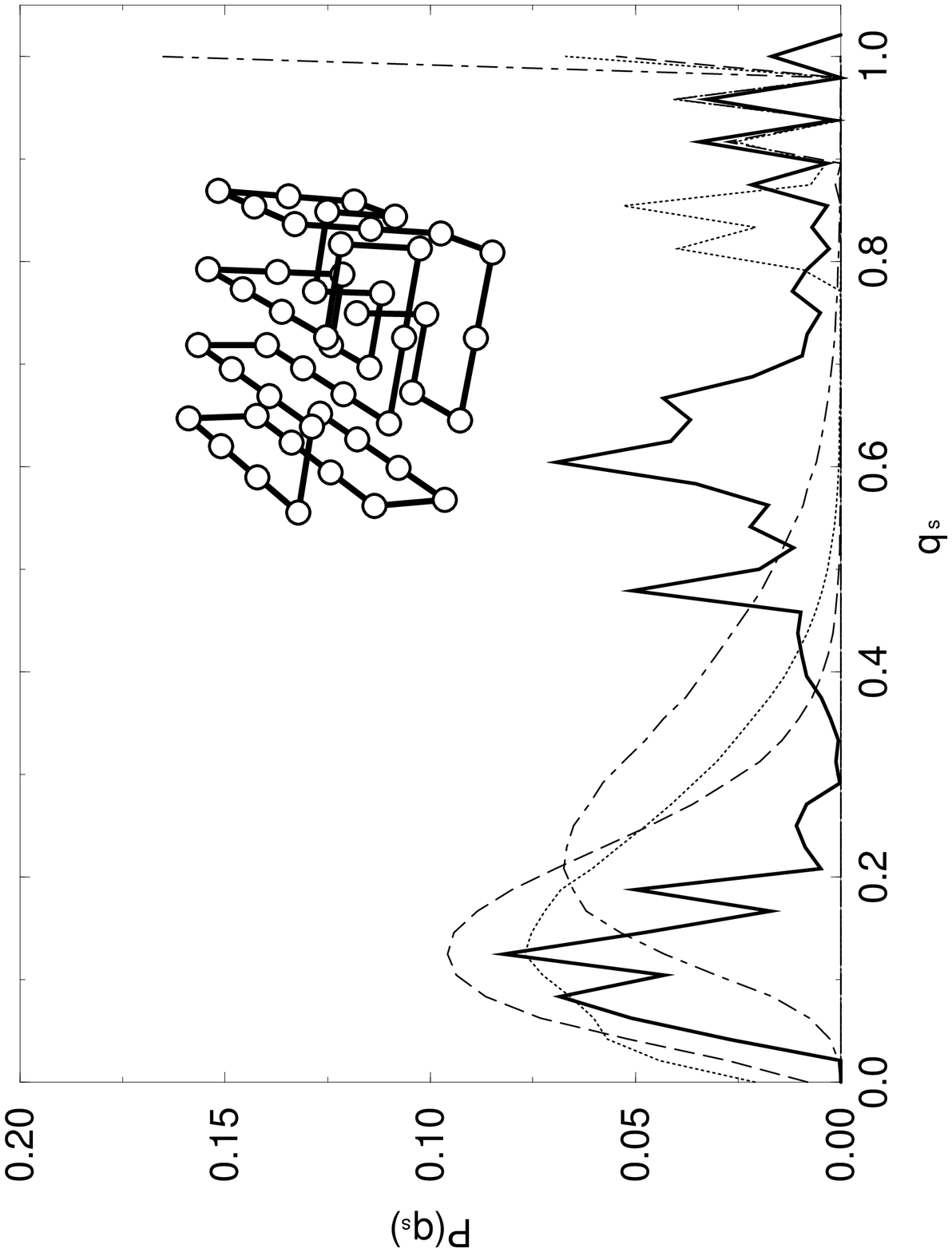}
\vspace{0.5in}

\pagebreak
\epsfysize 6 in
\epsffile{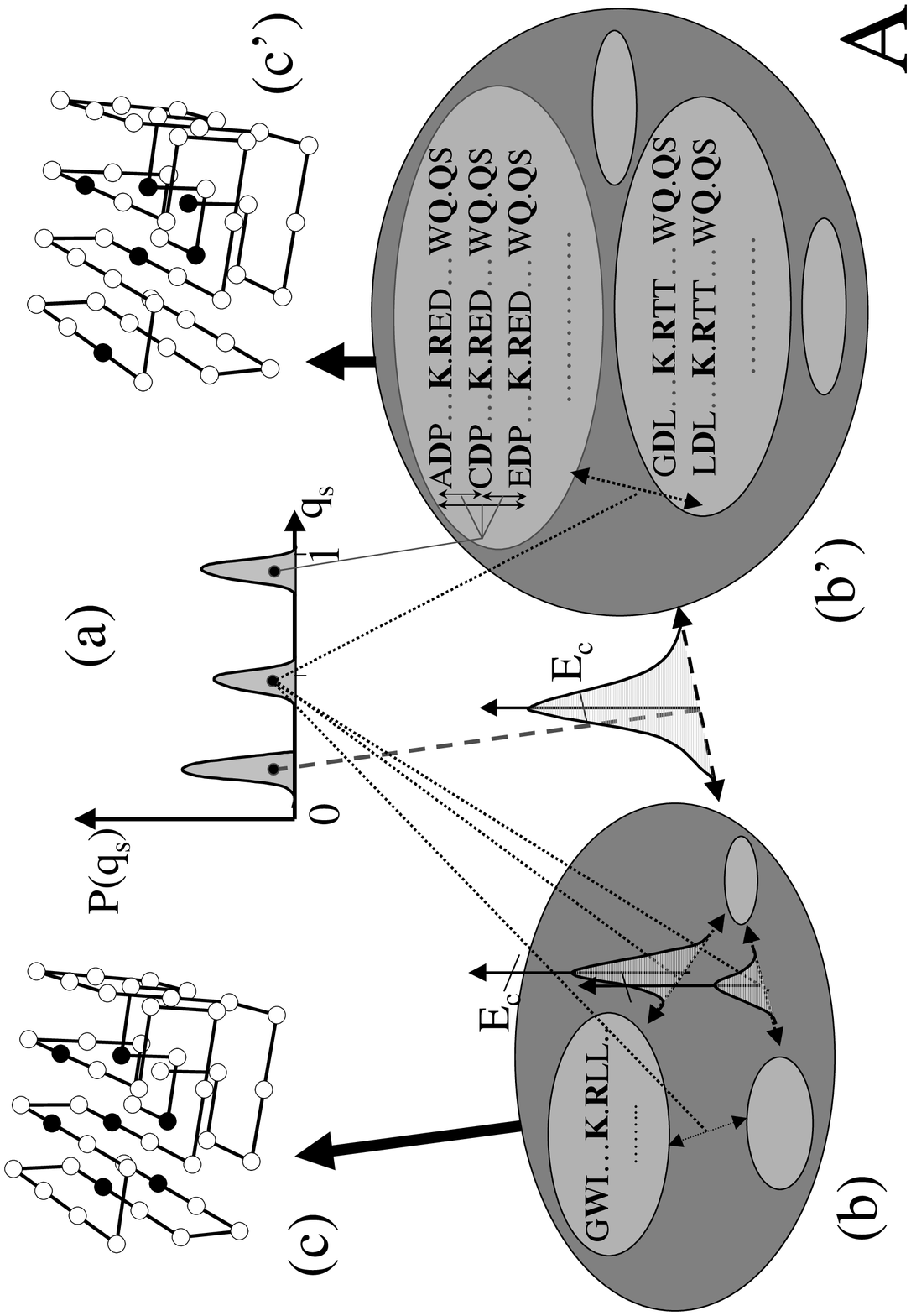}
\vspace{0.5in}


\pagebreak
\epsfysize 7 in
\epsffile{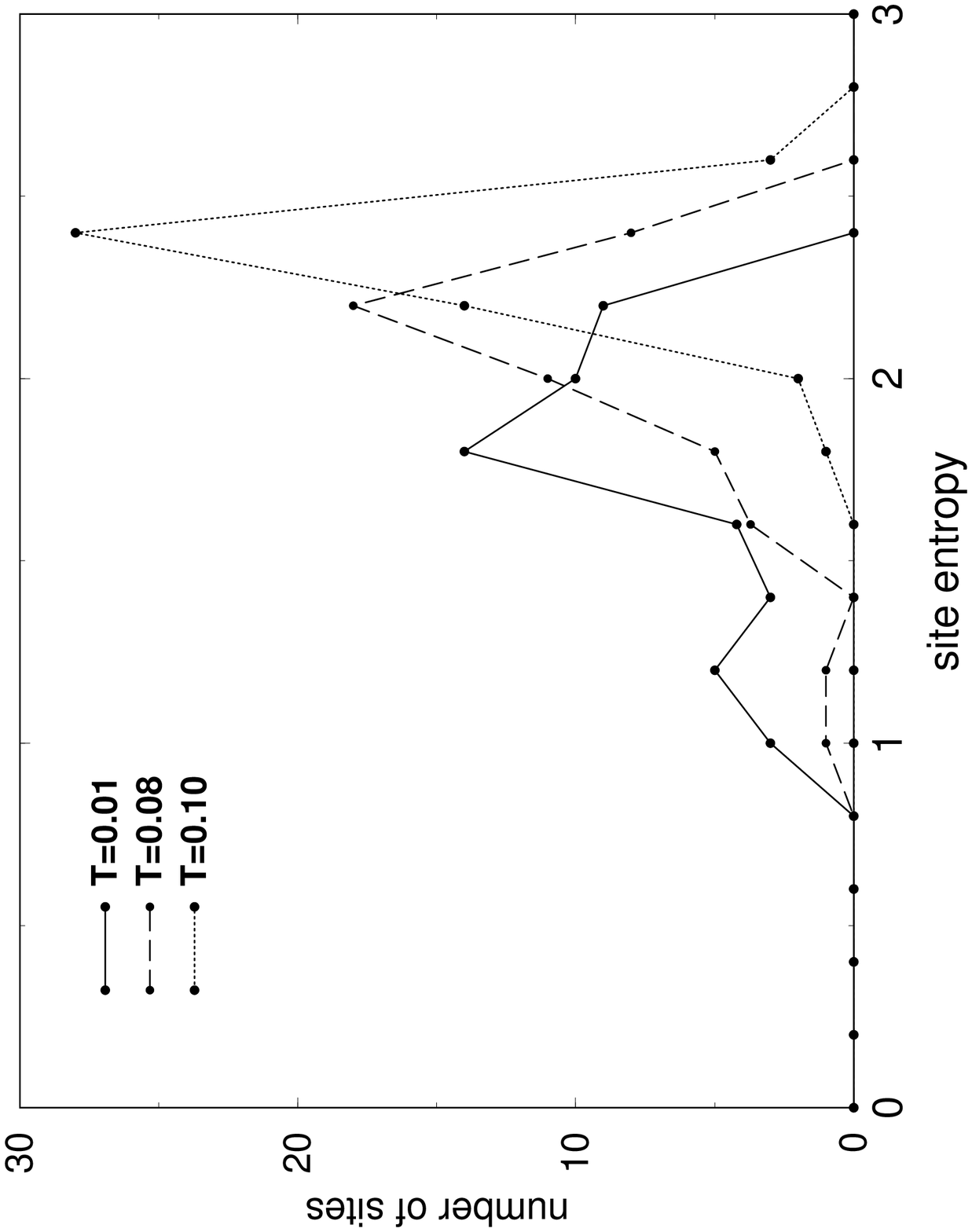}
\vspace{0.5in}

\pagebreak
\epsfysize 7 in
\epsffile{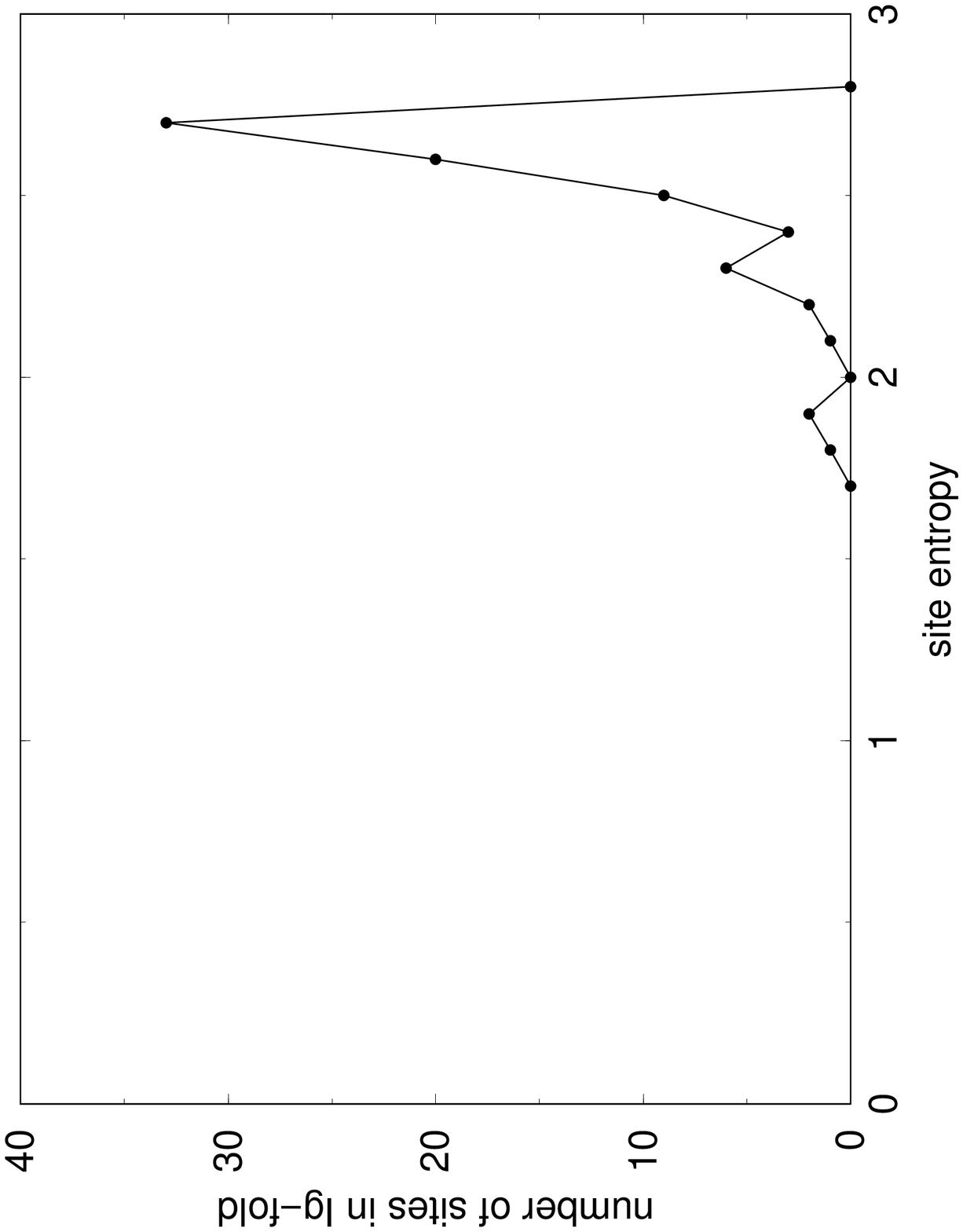}
\vspace{0.5in}

\end{document}